\documentclass{pasj00}

\draft

\begin{document}
\SetRunningHead{S. Kato}{}
\Received{2010/00/00}
\Accepted{2010/00/00}

\title{Trapped, Two-Armed, Nearly Vertical Oscillations in Disks with
        Toroidal Magnetic Fields}

\author{Shoji \textsc{Kato}}
\affil{2-2-2 Shikanoda-Nishi, Ikoma-shi, Nara, 630-0114}
\email{kato@gmail.com, kato@kusastro.kyoto-u.ac.jp}

%


\maketitle

\begin{abstract}
We have examined trapping of two-armed ($m=2$) nearly vertical oscillations
(vertical p-mode) in vertically isothermal ($c_{\rm s}=$ const.) relativistic
disks with toroidal magnetic fields.
The magnetic fields are stratified so that the Alfv{\'e}n speed, $c_{\rm A}$,
is constant in the vertical direction.
The ratio of $c_{\rm A}^2/c_{\rm s}^2$ in the vertical direction is taken
as a parameter examining the effects of magnetic fields on wave trapping.
We find that the two-armed nearly vertical oscillations are trapped in the 
inner region of disks and their frequencies decrease with increase of 
$c_{\rm A}^2/c_{\rm s}^2$.
The trapped regions of the fundamental ($n=1$) and the first-overtone ($n=2$)
are narrow (less than the length of the Schwarzschild radius, $r_{\rm g}$) and
their frequencies are relatively high (on the order of the angular frequency
of disk rotation in the inner region).
On contrast to this, the second-overtone ($n=3$) are trapped in a wide region
(a few times $r_{\rm g}$), and their frequencies are low and tend to zero
in the limit of $c_{\rm A}^2/c_{\rm s}^2=2.0$.
\end{abstract}

\section{Introduction}

Discoseismology is one of important fields in studying astrophysical disks, since
in some of them, quasi-periodic oscillations (QPOs) have been
observed and most of them seem to be attributed to disk oscillations.
In low mass X-ray binaries where the central sources are neutron stars or black
holes, for example, QPOs are often observed.
They are classified into high-frequency QPOs ($\geq 100$ Hz), low-frequency QPOs and low-frequency
complex (0.01 $\sim$ 100 Hz), power-law components, and others (van der Klis 2004).
The high-frequency and low-frequency QPOs will be, more or less, related to oscillation phenomena 
in the inner region of relativistic disks.

A first step to examine whether a disk oscillation mode can describe the observed
QPOs is to compare the frequencies resulting from the mode with those
of observed QPOs.
In disks which extend far outside, discrete frequencies of oscillations will be expected only when
oscillatory pertubations are trapped in a particular finite region of disks.
In this sense, examination of trapping of disk oscillations and of their 
frequencies is an important subject in discoseismology.

In geometrically thin relativistic disks, there are many kinds of disk oscillation modes.
Here, we classify them into four classes in terms of node numbers in the
vertical direction and the frequencies in the corotating frame, 
i.e., p-, g-, c-, and vertical p-mode oscillations (for details, see Kato et al. 2008;
Kato 2001).
(i) The p-mode is an inertial-acoustic mode with no node in the vertical direction\footnote{
When we mention the node number in the vertical direction, it is the number of node of density
perturbation in the vertical direction.
The node number associated with vertical component of velocity is smaller than that of density
perturbation by one.}
(nearly horizotal oscillations).
The square of their frequencies in corotationg frame, $(\omega-m\Omega)^2$, is larger 
than the square of the horizontal epicyclic frequency, $\kappa^2$, i.e.,
$(\omega-m\Omega)^2> \kappa^2$, where $\omega$ is the 
frequency of oscillation in the inertial frame, $m$ is the azimuthal wavenumber and
$\Omega$ is the angular velocity of disk rotation.
(ii) The g-mode oscillation (inertial mode or r-mode) has at least one node in the vetical
direction ($n\geq 1$) and ($\omega-m\Omega)^2$ is smaller than $\kappa^2$, i.e.,
$(\omega-m\Omega)^2<\kappa^2$.
(iii) The vertical p-mode oscillations are those that have at least one node in the vertical
direcin ($n\geq 1$) and have high frequencies in the corotating frame in the sense that 
$(\omega-m\Omega)^2>\Omega_\bot^2$, where $\Omega_\bot$ is the vertical epicyclic frequency and
always larger than $\kappa$.
This mode corresponds to "breathing mode" of Blaes et al. (2006) in oscillations of
tori.
(iv) Among oscillations formally belonging to $(\omega-m\Omega)^2>\Omega_\bot^2$, 
the one-armed ($m=1$) one with one node ($n=1$) in the vertical direction has a particular position.
The oscillation is called c-mode (corrugation mode).
It is nearly incompressible motions changing the disk plane up and down with a corrugation pattern,
corresponding to warp (or tilt).

Trapping of p-, g-, and c-mode oscillations have been extensively studied, e.g., by
Kato and Fukue (1980) and Ortega-Rodriguez et al. (2002) for p-mode oscillations; 
by Okazaki et al. (1987), Nowak and Wagoner (1992) and Perez et al. (1997) for g-mode 
oscillations; and by 
Kato (1990) and Silbergleit et al. (2001) for c-mode oscillations.
For reviews, see Wagoner (1999) and Kato (2001).
Recently, much development has been made in examination of effects of corotation resonance 
and magnetic fields on trapping and excitation (or damping) of oscillations.
That is, amplification of non-axisymmetric p-mode oscillations by corotation resonance was found by
Lai and Tsang (2009) and Tsang and Lai (2009b), which will be a refinding of the Papaloizou-Pringle
instability in different situations.
Different from the p-mode, non-axisymmetric g-mode and c-mode are havily damped by corotation
resonance (Kato 2003, Li et al. 2003, Latter and Balbus 2009 for g-mode, and Tsang and Lai 2009b
for c-mode).
Magnetic fields also have non-negligible effects on trapping.
Fu and Lai (2009) showed that 
magnetic fields act so as to destroy the general
relativistic self-trapping of axisymmetric g-mode oscillations.
Considering these recent developments, Lai and Tsang (2009)
suggested that among discoseismic modes a possible candidate of high-frequency QPOs will 
be axisymmetric p-mode oscillations.

Compared with the above many studies on p-, g- and c-mode oscillations, trapping
of vertical p-mode oscillations are little examined.
Examination of trapped vertical p-mode oscillations is, however, interesting
in relation to QPOs, since their frequencies are in a wide range by differences of 
i) node numbers in the vertical and radial directions, and
ii) disk parameters.
Furthermore, the general relativity is not essential in the trapping of vertical p-mode
oscillations.
Hence, the nearly vertical oscillations will be one of good candidates to describe 
the QPOs in various disks (disks of low-mass X-ray binaries to those of dwarf-navae)
by a unified model.

It should be noted that the interesting oscillations of vertical p-mode will be only those of
two-armed ones, i.e., $m=2$.
This is because there will be no trapping in oscillations of $m=1$, and because for oscillations of
$m\geq 3$ the frequencies of trapped oscillations are too high except for cases of large $n$.
The oscillations with large $n$, however, are not interesting from the observational points of view.
Considering these situations we restrict our attention only to two-armed ($m=2$) oscillations
with $n=1$ to $n=3$.

In previous paper (Kato 2010, paper I), we have examined trapping of two-armed vertical p-mode
oscillations in disks with polytopic gas, and showed how frequencies of trapped oscillations depend
on a change of polytropic index.
In this paper we restrict our attention to vertically isothermal disks for simplicity, 
but assume that the disks are subject to toroidal magnetic fields and examine how 
the frequency of trapped oscillations and the trapped region depend on the strength 
of magnetic fields.

\section{Unperturbed Disks and Equations Describing Disk Oscillations}

We consider geometrically thin, relativistic disks.
For mathematical simplicity, however, the effects of general relativity are taken into 
account only when we cosider radial distributions of $\Omega(r)$, $\kappa(r)$, and $\Omega_\bot(r)$, 
which are, in turn, the angular velocity of disk rotation, the epicyclic frequencies 
in the radial and vertical directions. 
Except for them, the Newtonial formulations are adopted.
Since geometrically thin disks are considered, $\Omega$, is approximated to be the relativistic 
Keplerian angular velocity, $\Omega_{\rm K}(r)$, when its numerical values are necessary.
Here, $r$ is the radial coordinate of cylindrical ones ($r$,$\varphi$,$z$), where the $z$-axis is
perpendicular to the disk plane and its origin is the disk center.
Functional forms of $\Omega_{\rm K}(r)$, $\kappa(r)$, and $\Omega_\bot(r)$ are given 
in many literatures (e.g., Kato et al. 2008).

\subsection{Unperturbed Disks with Toroidal Magnetic Fields}

The equilibrium disks are axisymmetric with toroidal magnetic fields.
The fields are assumed to be purely toroidal with no poloidal component:
\begin{equation}
      \mbox{\boldmath $B$}_0(r,z)=[0,B_0(r,z),0].
\label{2.1}
\end{equation}
We further assume that the gas is isothermal in the vertical direction and the magnetic fields 
$B_0$ are distributed in such a way that the Alfv\'{e}n speed $c_{\rm A}$ 
is constant in the vertical direction, i.e., $(B_0^2/4\pi\rho_0)^{1/2}=$ const. in the vertical
direction, where $\rho_0(r,z)$ 
is the density in the unperturbed disks.
Furthermore, the rotation is assumed to be cylindrical; e.g., the angular velocity of disk rotation, 
$\Omega$, is only a function of $r$.

We now consider the vertical structure of the disks.
The hydrostatic balance in the vertical direction is given by
\begin{equation}
     -\frac{1}{\rho_0}\frac{\partial}{\partial z}\biggr(p_0+\frac{B_0^2}{8\pi}\biggr)
         -\Omega_\bot^2z=0,
\label{2.2}
\end{equation}
where $p_0(r,z)$ is the pressure and related to $\rho_0(r,z)$ by $p_0=\rho_0c_{\rm s}^2$, 
$c_{\rm s}^2(r)$ being the isothermal acoustic speed.

Since both $c_{\rm s}$ and $c_{\rm A}$ are constant in the vertical direction, 
equation (\ref{2.2})
can be integrated to give
\begin{equation}
    \rho_0(r,z)=\rho_{00}(r){\rm exp}\biggr(-\frac{z^2}{2H^2}\biggr), \quad{\rm and}\quad
    B_0(r,z)=B_{00}(r){\rm exp}\biggr(-\frac{z^2}{4H^2}\biggr),
\label{2.3}
\end{equation}
where the scale height $H$ is related to $c_{\rm s}$, $c_{\rm A}$, and $\Omega_\bot$ by
\begin{equation}
      H^2(r)=\frac{c_{\rm s}^2+c_{\rm A}^2/2}{\Omega_\bot^2}.
\label{2.4}
\end{equation}

\subsection{Equations Describing Disk Oscillations}

Now, small-amplitude perturbations are superposed on the equilibrium disk described above.
The velocity perturbation over rotation is denoted by ($u_r$, $u_\varphi$, $u_z$), and
the perturbed part of magnetic field over the unperturbed one by ($b_r$, $b_\varphi$, $b_z$).
Then, the $r$-, $\varphi$-, and $z$-components of the linearized equation of motions are 
written, respectively, as
\begin{eqnarray}
   &&\biggr(\frac{\partial}{\partial t}+\Omega\frac{\partial}{\partial\varphi}\biggr)u_r
         -2\Omega u_\varphi  \nonumber \\
   && \hspace{30pt}  =-\frac{1}{\rho_0}\frac{\partial}{\partial r}\biggr(p_1+\frac{B_0b_\varphi}{4\pi}\biggr)
     +\frac{B_0}{4\pi \rho_0}\biggr(\frac{\partial b_r}{r\partial \varphi}-\frac{2b_\varphi}{r}\biggr)
     +\frac{\rho_1}{\rho_0^2}\biggr[\frac{\partial}{\partial r}\biggr(p_0+\frac{B_0^2}{8\pi}\biggr)
     +\frac{B_0^2}{4\pi r}\biggr],
\label{fluidr}
\end{eqnarray}
\begin{eqnarray}
   &&\biggr(\frac{\partial}{\partial t}+\Omega\frac{\partial}{\partial\varphi}\biggr)u_\varphi
    +\frac{\kappa^2}{2\Omega}u_r   \nonumber    \\
   &&\hspace{30pt}  =-\frac{1}{\rho_0}\frac{\partial}{r\partial\varphi}\biggr(p_1+\frac{B_0b_\varphi}{4\pi}\biggr)
    +\frac{B_0}{4\pi\rho_0}\biggr(\frac{\partial b_\varphi}{r\partial \varphi}+\frac{b_r}{r}\biggr)
    +\frac{1}{4\pi\rho_0}\biggr(b_r\frac{\partial}{\partial r}+b_z\frac{\partial}{\partial z}\biggr)B_0,
\label{fluidvarphi}
\end{eqnarray}
\begin{eqnarray}
   &&\biggr(\frac{\partial}{\partial t}+\Omega\frac{\partial}{\partial\varphi}\biggr)u_z  \nonumber  \\
   && \hspace{30pt} =-\frac{1}{\rho_0}\frac{\partial}{\partial z}\biggr(p_1+\frac{B_0b_\varphi}{4\pi}\biggr)
     +\frac{B_0}{4\pi \rho_0}\frac{\partial b_z}{r\partial \varphi}
     +\frac{\rho_1}{\rho_0^2}\frac{\partial}{\partial z}\biggr( p_0+\frac{B_0^2}{8\pi}\biggr),
\label{fluidz}
\end{eqnarray}
where $p_1$ and $\rho_1$ denote the perturbed parts of the pressure and density, respectively.
Similarly, the induction equation gives
\begin{equation}
   \biggr(\frac{\partial}{\partial t}+\Omega\frac{\partial}{\partial\varphi}\biggr)b_r
    =B_0\frac{\partial u_r}{r\partial\varphi},
\label{magr}
\end{equation}
\begin{equation}
   \biggr(\frac{\partial}{\partial t}+\Omega\frac{\partial}{\partial\varphi}\biggr)b_\varphi
    =r\frac{d\Omega}{dr}b_r-\frac{\partial}{\partial r}(B_0 u_r)
     -\frac{\partial}{\partial z}(B_0u_z),
\label{magvarphi}
\end{equation}
\begin{equation}
   \biggr(\frac{\partial}{\partial t}+\Omega\frac{\partial}{\partial\varphi}\biggr)b_z
    =B_0\frac{\partial u_z}{r\partial \varphi}.
\label{magz}
\end{equation}
The equation of continuity is
\begin{equation}
   \biggr(\frac{\partial}{\partial t}+\Omega\frac{\partial}{\partial\varphi}\biggr)\rho_1
    +\frac{\partial}{r\partial r}(r\rho_0u_r)
    +\frac{\partial}{r\partial\varphi}(\rho_0u_\varphi)
    +\frac{\partial}{\partial z}(\rho_0 u_z)=0.
\label{continuity}
\end{equation}
Another relation that we need here is a relation between $p_1$ and $\rho_1$.
Considering isothermal perturbations, we adopt 
\begin{equation}
       p_1=\rho_1c_{\rm s}^2.
\label{isothermal}
\end{equation}

Here, the azimuthal and time dependences of the perturbed quantities are taken to be proportional
to ${\rm exp}[i(\omega t-m\varphi)]$, where $\omega$ and $m$ are the frequency and
the azimuthal wavenumber of the perturbations, respectively.
The perturbations are assumed to be local in the sense that their characteristic radial wavelength,
$\lambda$, is shorter than the characteristic radial scale of disks, $\lambda_{\rm D}$,
i.e., $\lambda<\lambda_{\rm D}$.
By using this approximation, we neglect such quantities as
$d{\rm ln}\rho_{00}/d{\rm ln}r$, $d{\rm ln}B_{00}/d{\rm ln}r$, $d{\rm ln}H/d{\rm ln}r$,
and $d{\rm ln}\Omega/d{\rm ln}r$, compared with terms of the order of $r/\lambda$.
Then, the $r$-, $\varphi$-, and $z$-components of equation of motion, equations (\ref{fluidr}) --
(\ref{fluidz}), are reduced to
\begin{equation}
   i(\omega-m\Omega)u_r-2\Omega u_\varphi=-\frac{\partial h_1}{\partial r}-
          c_{\rm A}^2\frac{\partial}{\partial r}\biggr(\frac{b_\varphi}{B_0}\biggr),
\label{fluidr1}
\end{equation}
\begin{equation}
    i(\omega-m\Omega)u_\varphi+\frac{\kappa^2}{2\Omega}u_r=0,
\label{fluidvarphi1}
\end{equation}
\begin{equation}
    i(\omega-m\Omega)u_z=-\biggr(\frac{\partial}{\partial z}+\frac{c_{\rm A}^2}{2c_{\rm s}^2}
        \frac{z}{H^2}\biggr)h_1
        -c_{\rm A}^2\biggr(\frac{\partial}{\partial z}-\frac{z}{H^2}\biggr)\biggr(\frac{b_\varphi}{B_0}\biggr)
        -i\frac{m}{r}c_{\rm A}^2\biggr(\frac{b_z}{B_0}\biggr).
\label{fluidz1}
\end{equation}
In the above equations, $h_1$ defined by $h_1=p_1/\rho_0=c_{\rm s}^2\rho_1/\rho_0$ has been
introduced.
Similarly, the $r$-, $\varphi$-, and $z$-components of induction equation, equations (\ref{magr})
-- (\ref{magz}), are reduced to
\begin{equation}
     i(\omega-m\Omega)\frac{b_r}{B_0}=-i\frac{m}{r}u_r,
\label{magr1}
\end{equation}
\begin{equation}
     i(\omega-m\Omega)\frac{b_\varphi}{B_0}=r\frac{d\Omega}{dr}\frac{b_r}{B_0}
            -\frac{\partial u_r}{\partial r}
            -\biggr(\frac{\partial}{\partial z}-\frac{z}{2H^2}\biggr)u_z,
\label{magvarphi1}
\end{equation}
\begin{equation}
     i(\omega-m\Omega)\frac{b_z}{B_0}=-i\frac{m}{r}u_z.
\label{magz1}
\end{equation}
Finally, the equation of continuity, equation (\ref{continuity}), is reduced to
\begin{equation}
     i(\omega-m\Omega)h_1=-c_{\rm s}^2\biggr[\frac{\partial u_r}{\partial r}
            +\biggr(\frac{\partial}{\partial z}-\frac{z}{H^2}\biggr)u_z\biggr].
\label{continuity1}
\end{equation}

Now, we further simplify equations (\ref{fluidz1}) and (\ref{magvarphi1}).
The last term, $-i(m/r)c_{\rm A}^2(b_z/B_0)$, of equation (\ref{fluidz1}) 
can be expressed in terms of $u_z$ by using equation (\ref{magz1}).
The result shows that the term of $-i(m/r)c_{\rm A}^2(b_z/B_0)$ is smaller than 
the left-hand term, $i(\omega-m\Omega)u_z$, by a factor of $c_{\rm A}^2/r^2\Omega^2$.
Considering this, we neglect the last term on the right-hand side of 
equation (\ref{fluidz1}).
Next, we consider equation (\ref{magvarphi1}).
The first term on the right-hand side, $r(d\Omega/dr)(b_r/B_0)$, is 
smaller than the second term, $-\partial u_r/\partial r$, by
a factor of $\lambda/r$, which can be shown by expressing $b_r$ in terms of $u_r$ by
using equation (\ref{magr1}).
Hence, we neglect the term in the following analyses.

After introducing the above approximations into equations (\ref{fluidz1}) and 
(\ref{magvarphi1}), we multiply $i(\omega-m\Omega)$ to equation
(\ref{fluidz1}) in order to express $h_1$ and $b_\varphi/B_0$ in equation (\ref{fluidz1})
in terms of $u_z$ and $u_r$ by using equation (\ref{continuity1}) and (\ref{magvarphi1}).
Then, after changing independent variables from ($r$, $z$) to ($r$, $\eta$), where $\eta$
is defined by $\eta=z/H$, we have
\begin{equation}
    \biggr[\frac{\partial^2}{\partial \eta^2}-\eta\frac{\partial}{\partial \eta}
       +\frac{(\omega-m\Omega)^2-\Omega^2_\bot}{c_{\rm s}^2+c_{\rm A}^2}H^2\biggr]u_z 
            +H\biggr[\frac{\partial}{\partial\eta}
            -\frac{c_{\rm A}^2/2}{c_{\rm s}^2+c_{\rm A}^2}\eta\biggr]\frac{\partial u_r}{\partial r}=0.
\label{wave}
\end{equation}
This is the basic wave equation to be solved in this paper.

\section{Nearly Vertical Oscillations}

Equation (\ref{wave}) is now solved by approximately decomposing into two
equations describing behaviors in vertical and radial directions.

\subsection{Nearly Vertical Oscillations}

As mentioned before, we are interested in nearly vertical oscillations (i.e., vertical p-mode).
The main terms in equation (\ref{wave}) are thus those of the first brackets, and
the terms of the second brackets are small perturbed quantities.\footnote{
In nearly vertical oscillations, we have $u_z\sim h_1/c_{\rm s}$ [equation (\ref{fluidz1})].
Since the radial component of equation of motion [equation (\ref{fluidr1})] shows
that $u_r\sim (1/\Omega\lambda)h_1$, we have $u_r\sim (H/\lambda)u_z$, i.e., $u_r$
is smaller than $u_z$ by a factor of $H/\lambda$.
Hence, the terms with the second brackets of equation (\ref{wave}) is smaller than the terms with 
the first brackets by a factor of $(H/\lambda)^2$.
}
Although the terms of the second brackets are small quantities, they are of importance
to determine the wave trapping, as is shown in subsequent two subsections.

First, we should notice that the quantity $(\omega-m\Omega)^2-\Omega_\bot^2$ depends weakly 
on radius $r$. 
Hence, in order to consider this weak $r$-dependence of $(\omega-m\Omega)^2-\Omega_\bot^2$
as a small perturbed quantity, the third term in the first brackets of equation (\ref{wave}) is
now expressed as
\begin{equation}
       \frac{(\omega-m\Omega)^2-\Omega_\bot^2}{c_{\rm s}^2+c_{\rm A}^2}H^2
        =\biggr[\frac{(\omega-m\Omega)^2-\Omega_\bot^2}{c_{\rm s}^2+c_{\rm A}^2}H^2\biggr]_{\rm c}
           +\epsilon(r),
\label{eigenvalue0}
\end{equation}
where the subscript c represents the value at capture radius, $r_{\rm c}$, 
which is the outer boundary of the propagation region of oscillations and will be determined later,
and $\epsilon (r)$ is a small quantities depending on $r$.
The magnitude of $\epsilon(r)$ is found from equation (\ref{eigenvalue0}) when $r_{\rm c}$ and $\omega$ 
are determined later by an eigen-value problem in the radial direction (see the final subsection of 
this section). 
If the term of $\epsilon(r)$ is transported to terms of small perturbations, in the lowest 
order of approximations, equation (\ref{wave}) is written in the form:
\begin{equation}
    \frac{\partial^2}{\partial \eta^2}u_z^{(0)}-\eta\frac{\partial}{\partial \eta}u_z^{(0)}
    +\biggr[\frac{(\omega-m\Omega)^2-\Omega^2_\bot}{c_{\rm s}^2+c_{\rm A}^2}H^2\biggr]_{\rm c}u_z^{(0)}
    = 0,
\label{zeroth}
\end{equation}
where the superscript (0) is attached to $u_z$ in order to emphasize that it is the quantity of 
the lowest order of approximations.
By imposing the boundary condition that $u_z^{(0)}$ does not grow exponentially at $z=\pm\infty$,
we find that the $z$-dependence of $u_z^{(0)}$ can be expressed by a Hermite polynomial and
the term in the large brackets in equation (\ref{zeroth}) is determined as the eigenvalue 
and found to be $n-1$, where $n$ is a positive integer (Okazaki et al. 1987).
That is, we have
\begin{equation}
       u_z^{(0)} = f(r)g^{(0)}(\eta), 
\label{zerothuz}
\end{equation}
where 
\begin{equation}
        g^{(0)}(\eta)= {\cal H}_{n-1}(\eta), \quad n=1,2,3...
\label{eigenfunction}
\end{equation}
and
\begin{equation}
       \biggr[\frac{(\omega-m\Omega)^2-\Omega^2_\bot}{c_{\rm s}^2+c_{\rm A}^2}H^2\biggr]_{\rm c}
       = n-1.
\label{eigenvalue}
\end{equation}

Here, it is noted that the eigenfunction is taken to be ${\cal H}_{n-1}$, not ${\cal H}_n$.
The reason is that in many previous studies $h_1$ is adopted as the dependent variable (not $u_z$) and
the $z$-dependence of $h_1$ is taken to be 
proportional to ${\cal H}_n$ (e.g., Okazaki et al. 1987).
The node number of $u_z$ in the vertical direction is usually smaller than that of $h_1$
by unity (e.g., see the equation of the $z$-component of euation of motion).
Considering this, we have adopted ${\cal H}_{n-1}(\eta)$ instead of ${\cal H}_n(\eta)$ for $u_z$.  
As shown in equation (\ref{zerothuz}), in the lowest order of approximations, 
$u_z^{(0)}(r,\eta)$ is expressed in a separable form with respect to $r$ and $\eta$.
The $r$-dependence of $u_z^{(0)}$ is free at this stage, which is denoted by $f(r)$ 
in equation (\ref{zerothuz}).
It will be determined later by solving an eigen-value problem in the radial direction.

Equation (\ref{eigenvalue}) can be rewritten in the form 
\begin{equation}
       \biggr(\frac{\omega-m\Omega}{\Omega_\bot}\biggr)_{\rm c}^2= \biggr[\frac{c_{\rm s}^2
          +c_{\rm A}^2}{c_{\rm s}^2+c_{\rm A}^2/2}\biggr]_{\rm c}(n-1)+1.
\label{eigenvalue1}
\end{equation}
In the limit of $c_{\rm A}^2=0$, this equation is reduced to 
$(\omega-m\Omega)_{\rm c}^2=n\Omega_{\bot{\rm c}}^2$, which is the expected result from the local
dispersion relation in isothermal disks, i.e.,
$[(\omega-m\Omega)^2-\kappa^2][(\omega-m\Omega)^2-n\Omega_\bot^2]
=c_{\rm s}^2k^2(\omega-m\Omega)^2$ (Okazaki, et al. 1987), where $k$ is the radial wavenumber
of oscillations.
This dispersion relation shows that in non-magnetized isothermal disks, 
the propagation region of the nearly vertical oscillations is described by
$(\omega-m\Omega)^2>n\Omega_\bot^2$.
This means that for oscillations with $\omega$, one of their propagation region is  
$\omega<m\Omega-n^{1/2}\Omega_\bot$.
That is, the outer boundary of the propagation region on the $\omega$ - $r$ plane is 
given by $\omega=m\Omega-n^{1/2}\Omega_\bot$.
As shown later, equation (\ref{eigenvalue1}) suggests that in the present magnetized disks,
the outer boundary of the propagation region is given by
\begin{equation}
      \omega=m\Omega-\biggr[\frac{c_{\rm s}^2+c_{\rm A}^2}{c_{\rm s}^2
             +c_{\rm A}^2/2}(n-1)+1\biggr]^{1/2}\Omega_\bot.
\label{omega}
\end{equation}
This is really the case as is shown later.
      
\subsection{Derivation of Equation Describing Radial Behavior}

Now, we proceed to take into account the deviation of the oscillations from purely vertical 
ones as perturbations.
We see soon that separation of $u_z$
into two functions of $r$ and $\eta$ is no longer valid.
Hence, we consider the effects of small perturbed quantities by 
introducing a weak $r$-dependence in $g$.
That is, $u_z$ is now written as
\begin{equation}
         u_z(r,\eta)= f(r)[g^{(0)}(\eta)+g^{(1)}(r,\eta)+...].
\label{eigenfunction1}
\end{equation}
Then, from equation (\ref{wave}) we obtain, as an equation describing $fg^{(1)}$,
\begin{equation}
      f(r)\biggr(\frac{\partial^2}{\partial \eta^2}
                 -\eta\frac{\partial}{\partial \eta}+n-1\biggr)g^{(1)}(r,\eta)
            =-\epsilon(r)f(r)g^{(0)}(\eta)
             -H\biggr[\frac{\partial}{\partial \eta}-\frac{c_{\rm A}^2/2}{c_{\rm s}^2+c_{\rm A}^2}\biggr]
            \frac{\partial u_r^{(0)}}{\partial r},
\label{perturbation}            
\end{equation}
where $u_r^{(0)}$ is the lowest order expression for $u_r(r,\eta)$.

The next subject is to solve equation (\ref{perturbation}).
To do so, $u_r^{(0)}$ is expressed in terms of $u_z^{(0)}[=f(r)g^{(0)}(\eta)]$.
First, eliminating $u_\varphi$ from equations (\ref{fluidr1}) and (\ref{fluidvarphi1}), we have
\begin{equation}
     [-(\omega-m\Omega)^2+\kappa^2]u_r^{(0)}=-i(\omega-m\Omega)
         \biggr[\frac{\partial h_1^{(0)}}{\partial r}+c_{\rm A}^2\frac{\partial}{\partial r}
              \biggr(\frac{b_\varphi^{(0)}}{B_0}\biggr)\biggr].
\label{eliminate-uvarphi}
\end{equation}
In the lowest order of approximations of nearly vertical oscillations, the term of
$\partial u_r/\partial r$ on the right-hand side of equation (\ref{continuity1}) can be 
neglected in evaluating $h_1$, compared with the term of $(\partial/\partial z-z/H^2)u_z$.
Hence, by using equation (\ref{continuity1}) we can express $\partial h_1^{(0)}/\partial r$ 
on the right-hand side of equation (\ref{eliminate-uvarphi}) directly by $u_z$.
Furthermore, the main term on the right-hand side of equation (\ref{magvarphi1}) 
is the term with $u_z^{(0)}$.
Hence, by using equation (\ref{magvarphi1}) the term of $\partial(b_\varphi^{(0)}/B)/\partial r$
on the right-hand side of equation (\ref{eliminate-uvarphi}) can be also expressed in terms
of $u_z^{(0)}$.
Consequently, in the lowest order approximations of nearly vertical oscillations,
$u_r^{(0)}$ can be expressed in terms of $u_z^{(0)}$ alone from equation (\ref{eliminate-uvarphi}).
After some manipulations we have finally
\begin{equation}
     u_r^{(0)}={\cal L}_{\rm s}\biggr(\frac{\partial}{\partial \eta}-\eta\biggr)u_z^{(0)}
               +{\cal L}_{\rm A}\biggr(\frac{\partial}{\partial z}-\frac{1}{2}\eta\biggr)u_z^{(0)},
\label{expression-ur}
\end{equation}
where ${\cal L}_{\rm s}$ and ${\cal L}_{\rm A}$ are operators defined by 
\begin{equation}
     {\cal L}_{\rm s}=\frac{c_{\rm s}^2/H}{-(\omega-m\Omega)^2+\kappa^2}
           \biggr[\frac{\partial}{\partial r}
           -\frac{\partial{\rm ln}(\omega-m\Omega)}{\partial r}\biggr]
\label{Ls}
\end{equation}
and
\begin{equation}
     {\cal L}_{\rm A}=\frac{c_{\rm A}^2/H}{-(\omega-m\Omega)^2+\kappa^2}
           \biggr[\frac{\partial}{\partial r}
           -\frac{\partial{\rm ln}(\omega-m\Omega)}{\partial r}\biggr].
\label{LA}
\end{equation}

Now, we return to equation (\ref{perturbation}).
The equation is an inhomogeneous equation with respect to $g^{(1)}(r,\eta)$.
The right-hand side of the equation is now expressed in terms of $fg^{(0)}$ by using 
equation (\ref{expression-ur}).
As is done in a standard perturbation method, $g^{(1)}(r,\eta)$ is now expressed in a series of 
orthogonal functions of the zeroth order equations as     
\begin{equation}
     g^{(1)}(r,\eta)=\sum_m C_m(r){\cal H}_m(\eta).
\label{g1}
\end{equation}
The quantity $\epsilon(r)$ is then obtained from the solvability condition of
equation (\ref{perturbation}), using the orthogonality of the Hermite polynomials, which is
\begin{eqnarray}
     &&f\biggr\langle {\cal H}_{n-1}^2(\eta)\biggr\rangle\epsilon(r) \nonumber \\
     && +H\frac{d}{dr}{\cal L}_{\rm s}(f)\biggr\langle{\cal H}_{n-1}
               \biggr(\frac{d}{d \eta}
                -\frac{1}{2}\frac{c_{\rm A}^2}{c_{\rm s}^2+c_{\rm A}^2}\biggr)
               \biggr(\frac{d}{d\eta}-\eta\biggr){\cal H}_{n-1}\biggr\rangle  \nonumber  \\
     && +H\frac{d}{dr}{\cal L}_{\rm A}(f)\biggr\langle{\cal H}_{n-1}
               \biggr(\frac{d}{d \eta}
                -\frac{1}{2}\frac{c_{\rm A}^2}{c_{\rm s}^2+c_{\rm A}^2}\biggr)
               \biggr(\frac{d}{d\eta}-\frac{1}{2}\eta\biggr){\cal H}_{n-1}\biggr\rangle=0,    
\label{}
\end{eqnarray}
where $\langle A(\eta)B(\eta)\rangle$ is the integration of $A(\eta)B(\eta)$ with respect to $\eta$
in the range of ($-\infty$, $\infty$) with the weight ${\rm exp}(-\eta^2/2)$.

This solvability condition leads to an ordinary differential equation of $f(r)$, when the 
integration with respect to $\eta$ is performed.
After some manipulation we can write the results in the form:
\begin{equation}
   n\biggr(c_{\rm s}^2+\frac{1}{2}c_{\rm A}^2\biggr)\frac{d}{dr}
   \biggr[\frac{\omega-m\Omega}{(\omega-m\Omega)^2-\kappa^2}\frac{d}{dr}
        \biggr(\frac{f(r)}{\omega-m\Omega}\biggr)\biggr]
        +\epsilon(r)f(r)=0.
\label{radial-eq1}
\end{equation}
In the previous studies (paper I) on nearly vertical oscillations in non-magnetized disks, 
we have adopted $h_1$ (not $u_z$) as a dependent variable.
To compare our present results 
with those in the previous ones, we introduce here a new variable ${\tilde f}$ defined by
${\tilde f}=f/(\omega-m\Omega)$.\footnote{
In the lowest order of approximations, the equation of continuity gives
$i(\omega-m\Omega)h_1+(c_{\rm s}^2/H)(\partial/\partial \eta-\eta)u_z=0$.
If $u_z$ is taken to be proportional to ${\cal H}_{n-1}(\eta)$, i.e., $u_z=f(r){\cal H}_{n-1}(\eta)$, 
the above continuity relation shows that $h_1$ has a component proportional to ${\cal H}_n(\eta)$,
i.e., $h_1=f_h(r){\cal H}_n(\eta)$ and $f(r)$ and $f_h(r)$ is related by
$i(\omega-m\Omega)f_h=(c_{\rm s}^2/H)f$.
}
Then, equation (\ref{radial-eq1}) is reduced to
\begin{equation}
     \frac{1}{\omega-m\Omega}\frac{d}{dr}\biggr[\frac{\omega-m\Omega}
        {(\omega-m\Omega)^2-\kappa^2}\frac{d {\tilde f}}{dr}\biggr]   
        +\frac{\epsilon}{n\Omega_\bot^2H^2}{\tilde f}=0.
\label{radial-eq2}
\end{equation}
In the limit of $c_{\rm A}^2=0$, this equation becomes formally the same as that used 
in paper I.\footnote{
In paper I, polytropic disks are considered.
Hence, even in the limit of $c_{\rm A}^2$, equation (\ref{radial-eq2}) does not become idential 
with equation (33) in paper.
}

\subsection{Radial Eigenvalue Problems}

Next, we solve equation (\ref{radial-eq2}) as an eigen-value problem to know where the oscillations
are trapped and how much the eigen-frequency of the trapped oscillations are.
The same WKB procudures as Silbergleit et al. (2001) used are adopted here (see also paper I).
That is, we introduce a new independent variable $\tau(r)$ defined by
\begin{equation}
     \tau(r)=\int_{r_{\rm i}}^r\frac{{\tilde \omega}^2(r')-\kappa^2(r')}{-{\tilde \omega}(r')}dr', 
           \quad \tau_{\rm c}\equiv\tau(r_{\rm c}),
\label{3.19}
\end{equation}
where ${\tilde \omega}$ is defined by ${\tilde \omega}=(\omega-m\Omega)$,
and $r_{\rm i}$ is the inner edge of disks where a boundary condition is imposed.
Then, equation (\ref{radial-eq2}) is written in the form:
\begin{equation}
      \frac{d^2{\tilde f}}{d\tau^2} +Q {\tilde f}=0,
\label{3.20}
\end{equation}
where
\begin{equation}
      Q(\tau)=\frac{{\tilde\omega}^2}{{\tilde\omega}^2-\kappa^2}
          \frac{\epsilon}{n\Omega_\bot^2H^2}.
\label{3.22}
\end{equation}

Equations (\ref{3.20}) and (\ref{3.22}) show that the propagation region of oscillations is the region 
where $Q>0$, which is the region of $\epsilon >0$.
The region of $\epsilon(r)>0$ is found to be inside of $r_{\rm c}$ from the following considerations.
Let us assume tentatively that $r_{\rm c}$ is at a certain radius, although it should
be determined after solving equation (\ref{3.20}).
Since we take so that equation (\ref{eigenvalue}) holds at $r_{\rm c}$, equation (\ref{eigenvalue0}) 
defining $\epsilon$ gives
\begin{equation}
   \omega=m\Omega-\biggr[\frac{c_{\rm s}^2+c_{\rm A}^2}{c_{\rm s}^2+c_{\rm A}^2/2}(n-1)+1+\epsilon\biggr]^{1/2}
         \Omega_\bot.
\label{star*}
\end{equation}
This equation gives the $\omega$ - $r$ relation for a given $\epsilon$.
The $\omega$ - $r$ relation for $\epsilon =0$ is equation (\ref{omega}) and is shown in figures 1 and 2.
The curve monotonically increases inwards as $r$ decreases (see figures 1 and 2).
This means that if $r$ decreases from $r_{\rm c}$ keeping $\epsilon=0$, the frequency given by
equation (\ref{omega}) becomes larger than the frequency determined by $r_{\rm c}$.
Hence, to satisfy equation (\ref{star*}) under keeping $\omega$ at the value determined by $r_{\rm c}$, 
we must take a positive $\epsilon$.
If $r$ increases from $r_{\rm c}$ keeping $\omega$, on the other hand, equation (\ref{star*}) 
can be satisfied by taking a negative $\epsilon$.
In summary, we have $\epsilon>0$ inside of $r_{\rm c}$, while $\epsilon<0$ outside of $r_{\rm c}$.\footnote{
It is noted, however, that in the case where the curve of $\omega=m\Omega-[(n-1)(c_{\rm s}^2+c_{\rm A}^2)/
(c_{\rm s}^2+0.5c_{\rm A}^2)+1]^{1/2}\Omega_\bot$ monotonically decreases inwards, the situations are
changed (this is realized, for example, in the case where $n=3$ and $c_{\rm A}^2/c_{\rm s}^2>2$).
That is, we have $\epsilon>0$ outside of $r_{\rm c}$ and $\epsilon<0$ inside of $r_{\rm c}$.
This means that a propagation region of waves is outside of $r_{\rm c}$ and the waves are not trapped.
}

We solved equation (\ref{3.20}) by a standard WKB method with relevant boundary conditions
(for details, see Silbergleit et al. 2001).
The WKB approximation shows that the solution of equation (\ref{3.20}) can be represented as
\begin{equation}
    {\tilde f}\propto Q^{-1/4}(\tau){\rm cos}\ [\Phi(\tau)-\Phi_{\rm c}]
\label{3.21}
\end{equation}
in the whole capture region $0<\tau < \tau_{\rm c}$, except small vicinities of its boundaries
of $\tau=0$ (i.e., $r=r_{\rm i}$) and $\tau=\tau_{\rm c}$ (i.e., $r_{\rm c}$).
Here, $\Phi(\tau)$ is defined by
\begin{equation}
     \Phi(\tau)=\int_0^\tau Q^{1/2}(\tau')d\tau'=\int_{r_{\rm i}}^rQ^{1/2}(r')
          \frac{{\tilde \omega}^2(r')-\kappa^2(r')}{-{\tilde\omega}(r')}dr',
\label{3.23}
\end{equation}
and $\Phi_{\rm c}$ is a constant to be determined by boundary conditions.
To determine the outer boundary condition, we take into account the fact that 
the capture radius, $r_{\rm c}$, 
is a turning point of equation (\ref{3.20}) since the sign of $\epsilon$ changes there.
The inner boundary condition we adopted is 
${\tilde f}=0$ or $d{\tilde f}/dr=0$ at $r_{\rm i}$.
As the inner boundary radius we take the marginary stable radius.
Then, WKB analyses show that the trapping condition is
\begin{eqnarray}
    \int_0^{\tau_{\rm c}}Q^{1/2}d\tau=\left\{\begin{array}{ll} \pi(n_r+1/4) 
                                    & {\rm for}\quad d{\tilde f}/dr=0 \\
                                                   \pi(n_r+3/4) 
                                    &{\rm for}\quad  {\tilde f}=0, 
                                                                 \label{3.24}    
                    \end{array}
      \right. 
\end{eqnarray}
where $n_r(=0,1,2,...)$ is zero or a positive integer specifying the node number of 
${\tilde f}$ in the radial direction.
The constant $\Phi_{\rm c}$ is determined as
\begin{eqnarray}
    \Phi_{\rm c}=\left\{\begin{array}{ll} 0 
                                    & {\rm for}\quad d{\tilde f}/dr=0 \\
                                         \pi/2
                                    &{\rm for}\quad  {\tilde f}=0. 
                                                                 \label{3.25}    
                    \end{array}
      \right. 
\end{eqnarray}

For a given set of parameters, including spin parameter $a_*$ 
and mass of neutron stars, $M$, any solution of equation (\ref{3.24}) specifies $r_{\rm c}$,
which gives $\omega$ of the trapped oscillation through equation (\ref{eigenvalue1}).
In other words, $\omega$ and $r_{\rm c}$ are related by equation (\ref{eigenvalue1}), i.e., 
$\omega=\omega(r_{\rm c})$ or $r_{\rm c}=r_{\rm c}(\omega)$, and
the trapping condition determines $r_{\rm c}$ or $\omega$ as functions of such 
parameters as $c_{\rm S}^2$, $c_{\rm A}^2$, $a_*$ and $M$.

\section{Numerical Results}

To obtain numerical values of the frequency, $\omega$, and the capture radius, $r_{\rm c}$, 
of trapped oscillations, we must
specify the radial distribution of acoustic speed, i.e., $c_{{\rm s}0}(r)$.
The final results of numerical calculations show that the trapped region is in the inner region of
the disks.  
Hence, we consider the temperature distribution in the standard disk where gas pressure dominates 
over radiation pressure and
opacity mainly comes from the free-free processes, and adopt (e.g., Kato et al. 2008) 
\begin{equation}
     c_{{\rm s}0}^2=1.83\times 10^{16}(\alpha m)^{-1/5}{\dot m}^{3/5}r^{-9/10}\ {\rm cm}^2\ {\rm s}^{-2},
\label{4.1}
\end{equation}
where $\alpha$ is the conventional viscosity parameter, $m(\equiv M/M_\odot)$\footnote{
In this section and hereafter, $m$ is often used to denote $M/M_\odot$ without confusion with the
azimuthal wavenumber $m$ of oscillations.
} 
and ${\dot m}={\dot M}/{\dot M}_{\rm crit}$, ${\dot M}_{\rm crit}$ being the critical mass-flow
rate defined by 
\begin{equation}
        {\dot M}_{\rm crit}\equiv\frac{L_{\rm E}}{c^2}=1.40\times 10^{17}m\ {\rm g}\ {\rm s}^{-1},
\label{4.2}
\end{equation}
where $L_{\rm E}$ is the Eddington luminosity.
Parameters $\alpha$ and ${\dot m}$ affect on the frequencies of trapped oscillations only 
through the magnitude of $c_{{\rm s}0}$.
We adopt, throughout this paper, $\alpha=0.1$ and ${\dot m}=0.3$.
A parameter specifying the strength of magnetic field is $c_{\rm A}^2/c_{\rm s}^2$.
In this paper, we consider the disks where the parameter $c_{\rm A}^2/c_{\rm s}^2$
is in the range of $c_{\rm A}^2/c_{\rm s}^2 = 0 \sim 2$.
Other parameters specifying the disk-star system are $m(\equiv M/M_\odot)$ and $a_*$.
We consider the cases of $M/M_\odot=2.0$ and $a_*=0\sim 0.3$.

We only consider two-armed oscillations with one, two, or three node(s) in the vertical direction,
i.e., $n=1$, 2, or 3.\footnote{
In oscillations with $n$, we have $h_1(r,z)\propto {\cal H}_n(z/H)$ and $u_z(r,z)\propto
{\cal H}_{n-1}(z/H)$.
That is, in oscillations with $n=3$, $u_z$ is plane-symmetric with respect to the equatorial plane,
and has one node (where $u_z=0$) above and below the equator.
}
Oscillations with more nodes in the vertical direction are less interesting from
the view point of observability.
The inner boundary of oscillations is taken at the radius of $\kappa=0$, i.e.,
at the radius of the marginally stable circular orbit.
In this paper, $u_z=0$ (i.e., ${\tilde f}=0$) is adopted at the radius as a boundary condition,
except in figure 5.
In figure 5, boundary condition of $du_z/dr$ (i.e., $d{\tilde f}/dr\sim 0$) is considered as well as 
$u_z=0$ in order to see effects of boundary condition on results.
We find that the differences of the boundary condition bring about quantitative differeces in
results, but there is no essential differences in parameter dependences of results.
Hence, except in figure 5, we adopt $u_z=0$ as the inner boundary condition throughout this paper.
The horizontal node number, $n_{\rm r}$, of oscillations we consider is mainly $n_{\rm r}=0$ and 
supplementally $n_{\rm r}=1$ and 2.

Figures 1 and 2 are the propagation diagrams for oscillations of $n=1$ and 2 (figure 1) and $n=3$
(figure 2), respectively, in the disks with $c_{\rm A}^2/c_{\rm s}^2=1$ and $a_*=0$.
Only the oscillations of $n_{\rm r}=0$ are shown in figure 1, but
three modes of oscillations, i.e., $n_{\rm r}=0$, 1, and 2, are shown in figure 2.
The propagation regions of oscillations on the frequency-radius diagram is
below the curve given by 
$\omega= m\Omega-[(n-1)(c_{\rm s}^2+c_{\rm A}^2)/(c_{\rm s}^2+c_{\rm A}^2/2)+1]^{1/2}
\Omega_\bot$ [see equation (\ref{omega})].
The results of numerical calculations show that the oscillations of $n=1$ with $n_{\rm r}=0$ are 
trapped in the radial range shown by the upper thick horizontal line in figure 1.
The frequency $\omega$ and the capture radius $r_{\rm c}$ are, respectively, $\omega=798$Hz and 
$r_{\rm c}=3.72r_{\rm g}$.
Outside $r_{\rm c}$, the oscillation is spatially damped.
The radial range of trapped oscillations of $n=2$ with $n_{\rm r}=0$  
is shown by the lower thick horizontal line in figure 1.
The frequency and the capture radius in this case are $\omega=359$Hz and $r_{\rm c}=3.84r_{\rm g}$.

Trapped oscillations of $n=3$ have frequencies lower than those of $n=1$ and 2,
since on the propagation diagram the curve of 
$\omega=m\Omega-[(n-1)(c_{\rm s}^2+c_{\rm A}^2)/(c_{\rm s}^2+c_{\rm A}^2/2)+1]^{1/2}
\Omega_\bot$
is below those in the cases of $n=1$ or $n=2$.
In figure 2, the frequency and the radial extend of trapped oscillations of $n=3$
are shown for three modes in the radial direction; 
the fundamental mode (i.e., $n_{\rm r}=0$) and the first two overtones
(i.e., $n_{\rm r}=1$ and 2).
The sets of frequency and capture radius for these three modes of $n_{\rm r}=0$, 1, and 2 are, respectively,
(49.5Hz, 4.60$r_{\rm g}$), (28.4Hz, 6.66$r_{\rm g}$), and (17.2Hz, 9.31$r_{\rm g}$)
in the disks with $c_{\rm A}^2/c_{\rm s}^2=1.0$ and $a_*=0$.

Figure 3 shows the $c_{\rm A}^2/c_{\rm s}^2$-dependence of the capture radius $r_{\rm c}$.
As a typical case, the dependence is shown for oscillations of $n_{\rm r}=0$.
No spin of the central source is adopted.
It is noted that when $c_{\rm A}^2/c_{\rm s}^2$ is close to 2, the capture radius of oscillations 
of $n=3$ is far outside and the frequencies are low.
These characteristics become more prominent for oscillations with $n_{\rm r}\geq 1$, although
they are not shown in figure 3 (see figure 5).
As $c_{\rm A}^2/c_{\rm s}^2$ increases beyond 2, the oscillations of $n=3$ are no longer 
trapped.
This is related to the behavior of the curve of 
$\omega=m\Omega-[(n-1)(c_{\rm s}^2+c_{\rm A}^2)/(c_{\rm s}^2+c_{\rm A}^2/2)+1]^{1/2}
\Omega_\bot$ on the $\omega$ - $r$ plane.
If $n=3$ and $c_{\rm A}^2/c_{\rm s}^2>2$, $\omega$ given by the above relation is negative and  
is a monotonically increasing function outwards on the $\omega$ - $r$ plane.
Then, the region of $\epsilon>0$ (i.e., propagation region) is in the outer region and there is no 
trapping (see the previous section).

The frequency-$c_{\rm A}^2/c_{\rm s}^2$ relations are summarized 
in figure 4 for two disks with $a_*=0$ and $a_*=0.2$.
Modes of oscillations adopted are $n=1$, 2, and 3.
In all cases $n_{\rm r}$ is taken to be $n_{\rm r}=0$.
As mentioned before, the oscillations with $n=3$ have low frequencies.
In order to examine characteristics of these low frequency oscillations more in detail, 
the frequency-$c_{\rm A}^2/c_{\rm s}^2$ relation in case of $n=3$ is again shown in figure 5, 
including cases where other parameter values are adopted.
That is, in addition to oscillations with $n_{\rm r}=0$, oscillations with $n_{\rm r}=1$ and 2
are considered in figure 5.
In addition, the cases where $(d{\tilde u}_z)/dr=0$ is adopted at $r_{\rm i}$
as the inner boundary condition are shown by thin curves.
In figure 6, the frequency - spin relation is shown for three modes of oscillations with
$n=1$ , 2, and 3, where $n_{\rm r}=0$ and $c_{\rm A}^2/c_{\rm s}^2=1.0$ are adopted.

\begin{figure}
\begin{center}
    \FigureFile(80mm,80mm){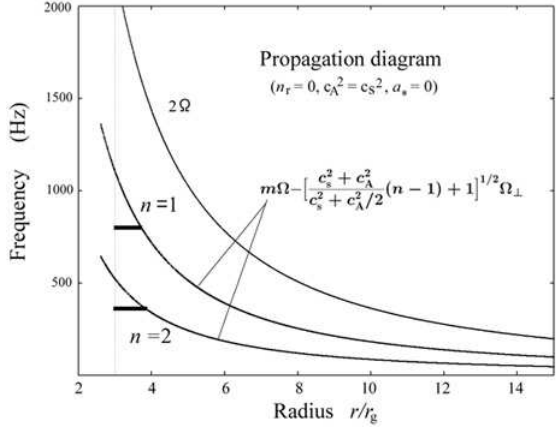}
\end{center}
\caption{
Frequency - radius plane (i.e., propagation diagram) showing the propagation region of two-armed 
($m=2$) nearly vertical oscillations (i.e., vertical p-mode oscillations) in vertically isothermal disks
with toroidal magnetic fields. 
The value of $c_{\rm A}^2/c_{\rm s}^2= 1.0$ has been adopted.
The propagation region of oscillation modes with $n$ is below the boundary curve 
labelled by $m\Omega-[(n-1)(c_{\rm s}^2+c_{\rm A}^2)/(c_{\rm s}^2+0.5c_{\rm A}^2)+1]^{1/2}\Omega_\bot$.
The boundary curve is shown for two cases of $n=1$ and $n=2$.
In the case of $n=1$ (and $n_{\rm r}=0$), the capture (trapped) zone and the frequency of the 
trapped oscillations are shown by the upper thick horizontal line 
(the frequency $\omega$ is $\sim$ 798Hz and capture radius $r_{\rm c}$ is $\sim 3.72r_{\rm g}$).
In oscillations with $n=2$, 
the trapped oscillation with $n_{\rm r}=0$ is shown by the lower thick horizontal line.
The trapped frequency $\omega$ is $\sim 360$Hz and $r_{\rm c}\sim 3.84 r_{\rm g}$.
The inner boundary condition adopted at $r_{\rm i}$ is ${\tilde f}=0$.
This inner boundary condition is adopted in all cases in this paper, except for in
figure 5.
The central star is assumed to have no spin.
The mass of the central star is taken to be $2M_\odot$ in all cases shown in figures in this paper.
} 
\end{figure}
\begin{figure}
\begin{center}
    \FigureFile(80mm,80mm){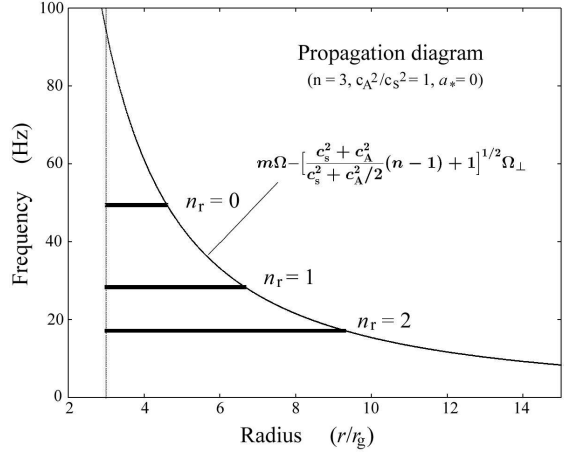}
\end{center}
\caption{
The same as figure 1, except that the oscillations with $n=3$ are considered here.
The propagation region of the oscillations is below the curve labelled by
$m\Omega-[(n-1)(c_{\rm s}^2+c_{\rm A}^2)/(c_{\rm s}^2+0.5c_{\rm A}^2)+1]^{1/2}\Omega_\bot$,
where $n=3$ is taken.
Trapping of three modes of oscillations with $n_{\rm r}=0$, 1, and 2 are shown by 
three horizontal thick lines.
The sets of frequency and capture radius for these three oscillation modes are, in turn, 
(49.5Hz, 4.60$r_{\rm g}$), (28.4Hz, 6.66$r_{\rm g}$), and (17.2Hz, 9.30$r_{\rm g}$).
It is noted that $c_{\rm A}^2/c_{\rm s}^2=1.0$ is adopted here, but in the case of 
$c_{\rm A}^2/c_{\rm s}^2=2.0$, there is no trapped oscillations.
} 
\end{figure}
\begin{figure}
\begin{center}
    \FigureFile(80mm,80mm){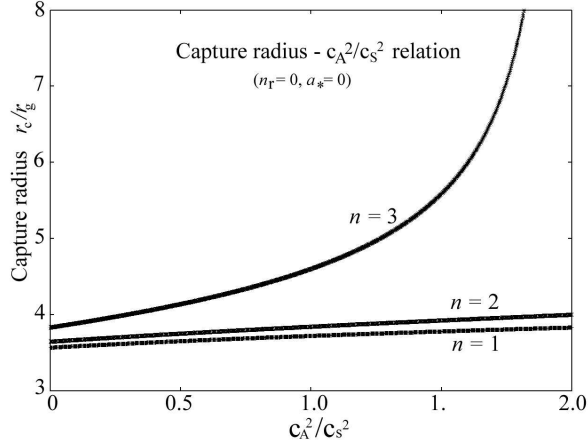}
\end{center}
\caption{
Capture radius, $r_{\rm c}$, as functions of $c_{\rm A}^2/c_{\rm s}^2$ for three modes of $n=1$, 2, 
and 3.
The radial node number, $n_{\rm r}$, is taken to be zero with boundary condition ${\tilde f}=0$ at 
$r_{\rm i}=3r_{\rm g}$, the spin parameter being $a_*=0$. 
In oscillations with $n=3$, the trapping zone extends infinity as $c_{\rm A}^2/c_{\rm s}^2$ approaches
$c_{\rm A}^2/c_{\rm s}^2=2$, and no trapping for $c_{\rm A}^2/c_{\rm s}^2>2$.
No spin of the central source is taken, i.e., $a_*=0$.   
} 
\end{figure}
\begin{figure}
\begin{center}
    \FigureFile(80mm,80mm){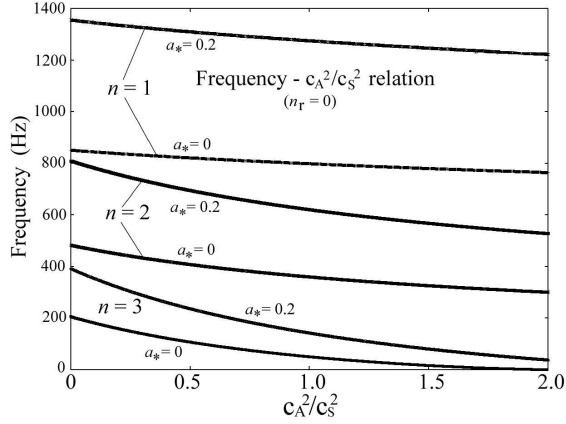}
\end{center}
\caption{
Frequency-$c_{\rm A}^2/c_{\rm s}^2$ relation of trapped oscillations for some values of 
vertical node number $n$ and spin parameter $a_*$.
The oscillations with no node in the radial direction ($n_{\rm r}=0$) are considered. 
} 
\end{figure}
\begin{figure}
\begin{center}
    \FigureFile(80mm,80mm){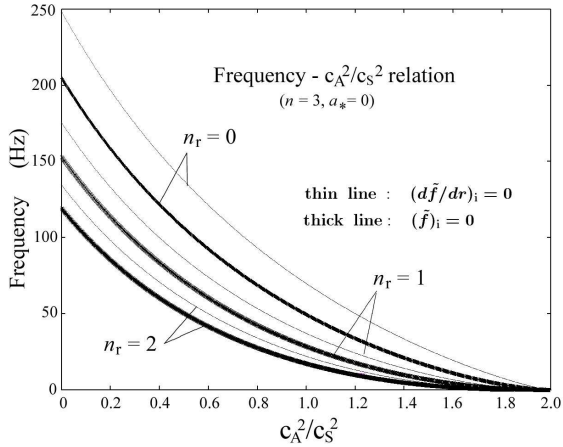}
\end{center}
\caption{
Frequency-$c_{\rm A}^2/c_{\rm s}^2$ relation for oscillation modes with $n=3$.
This figure is to demonstrate the effects of differences of radial node number $n_{\rm r}$ 
and of boundary condition on the frequency of trapped oscillations. 
Two cases of boundary conditions, ${\tilde f}_{\rm i}=0$ [which corresponds to $(h_1)_{\rm i}=0$]
and $(d{\tilde f}/dr)_{\rm i}=0$ [which corresponds to $(dh_1/dr)_{\rm i}=0$] 
are compared for three modes of oscillations with radial node number $n_{\rm r}=0$, 1, and 2.
The thick curves are for the cases where the inner boundary condition is taken
as ${\tilde f}_{\rm i}=0$,
while the thin curves are for the cases of $(d{\tilde f}/dr)_{\rm i}=0$.
The spin parameter $a_*$ is taken to be zero. 
} 
\end{figure}
\begin{figure}
\begin{center}
    \FigureFile(80mm,80mm){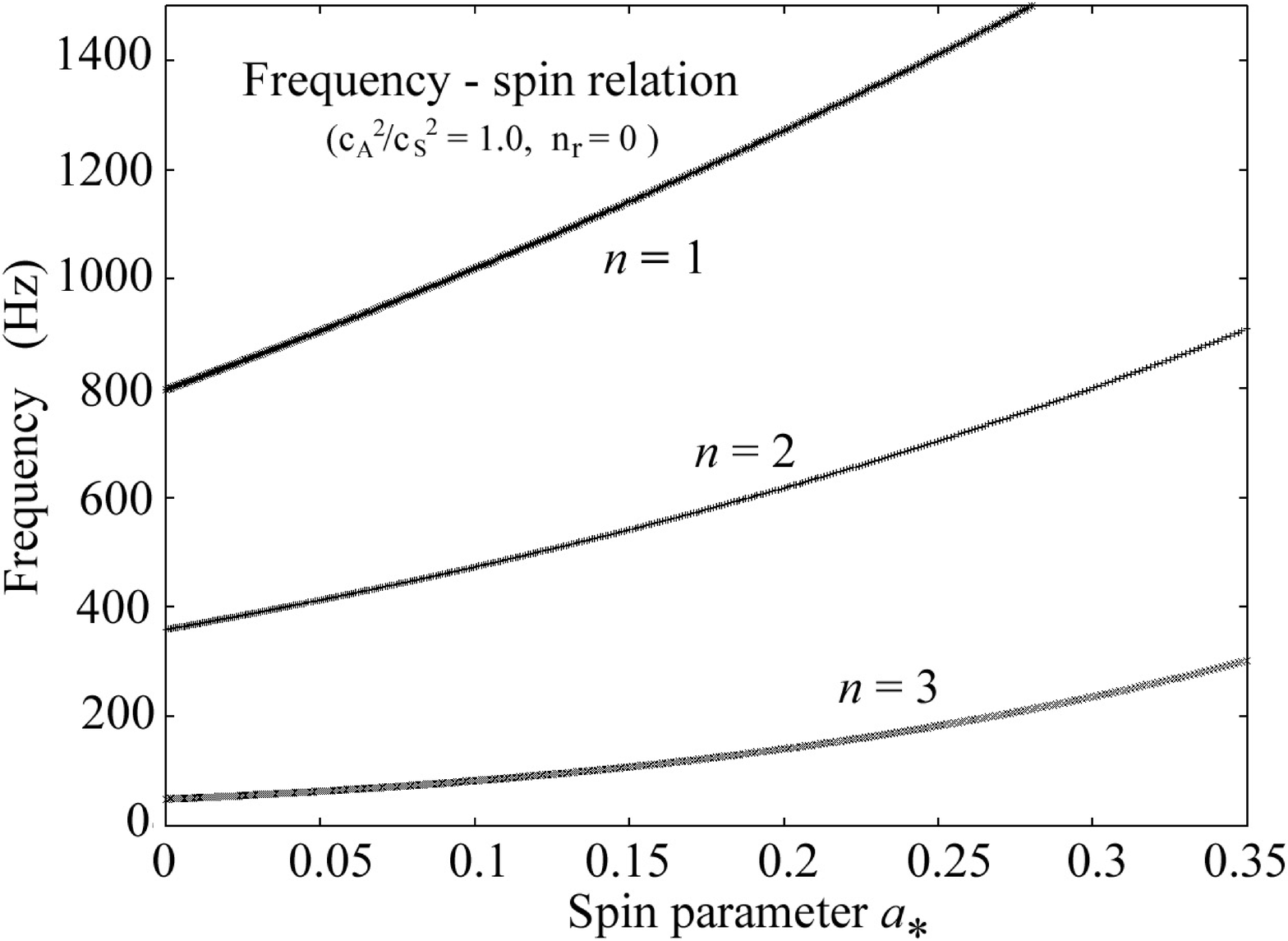}
\end{center}
\caption{
Frequency-spin relation for modes with $n=1$, 2, and 3.
The radial node number is taken to be $n_{\rm r}=0$,
$c_{\rm A}^2/c_{\rm s}^2=1.0$, and $a_*=0$ have been adopted.
} 
\end{figure}

\section{Discussion}

In this paper we have examined characteristics of trapping of two-armed ($m=2$), nearly vertical 
oscillations (vertical p-mode oscillations), assuming that the disk is isothermal in the vertical 
direction and is subject to purely toroidal magnetic fields.
Fot mathematical simplicity, the ratio of the Alfv{\' e}n speed to the acoustic apeed is constant 
in the vertical direction, i.e., $c_{\rm A}^2/c_{\rm s}^2$ is constant in the vertical direction.
The effects of magnetic fields on characteristics of the oscillation are measured by taking 
$c_{\rm A}^2/c_{\rm s}^2$ as a parameter.

A differece from paper I is that in paper I we have considered a polytropic gas 
(not an isothermal gas) and examined how the characteristics of trapping of the nearly vertical 
oscillations depend on the polytropic index characterizing the gas.
In this paper, the disk gas is taken to be isothermal, for simplicity, differet from the paper I, 
but is subject to toroidal magnetic fields.
Under these situations, we examine how the characteristics of nearly vertical
oscillations change by a change of strength of magnetic fields.

In this paper a partial differential equation [equation (\ref{wave})] has been solved by a 
perturbation method.
In this procedure, we have assumed that $\epsilon(r)$ defined by equation (\ref{eigenvalue0}) 
is a small positive quantitiy (i.e., $\epsilon <1$) in the wave propagation region.
The final results show that this is really acceptable as a first step to examine qualitative behavior 
of trappeing.
To do quantitative argument, however, the approximation should be improved especially in the case
of $n=1$ and $n=2$.
That is, $\epsilon(r)$ is zero at $r=r_{\rm c}$ by definition and increases inwards monotonically
and becomes a maximum at $r=r_{\rm i}$.
The results of calculations show that the maximum value of $\epsilon$ is 0.46 for $n=1$, 0.35 for
$n=2$, and 0.12 for $n=3$ when $c_{\rm A}^2/c_{\rm s}^2=1.0$, $a_*=0$ and $n_{\rm r}=0$.
The value slightly increases with decrease of $c_{\rm A}^2/c_{\rm s}^2$ and increase of
$a_*$ and $n_{\rm r}$.

In this paper we did not quantitatively consider the effects of $c_{\rm s0}(r)$ on frequency.
An increase of $c_{\rm s0}$ without any change of other parameters leads to decrease of 
frequency of trapped oscillations.
The reason is that an increase of $c_{\rm s0}$ decreases $Q$.
Hence, to satisfy the trapping condition (\ref{3.24}), an increase of $r_{\rm c}$ is necessary,
which leads to decrease of frequency (see figures 1 and 2).
The effects of changes of other various parameter values on frequencies of trapped oscillations are 
qualitatively the same as those in paper I (see table1 in paper I).

The purpose of this paper is to demonstrate the importance of two-armed ($m=2$), nearly
vertical oscillations as one of possible candidates of
disk oscillations describing quasi-periodic oscillations observed in
low-mass X-ray binaries (LMXBs).
One of reasons why we take attention on these oscillations is that they can be trapped in the inner
region of disks and their frequencies can cover a wide range of frequency by
(i) differece of modes ($n$ and $n_{\rm r}$), and (ii) change of disk structure.
As a change of disk structure, we considered a change of polytropic index in paper I, 
while we consider here a change of magnetic fields.

The parameters describing the difference of oscillation modes are $n(=1,2,3,...$) and
$n_{\rm r}(=0,1,2,...)$, where $n$ and $n_{\rm r}$ are, respectively, the node number of 
$h_1$ in the vertical and horizontal directions.
As in paper I, the trapped oscillations of $n=1$ and $n=2$ have frequencies on the order of
kHz QPOs, while those of $n=3$ have lower frequencies and on the order of horizontal branch 
and normal branch oscillations.
An interesting result obtained in this paper is that the frequency of trapped oscillations decreases as 
magnetic fields become stronger (figures 4 and 5).
This is particularly so for oscillations of $n=3$ (figure 5).
That is, the trapped region of oscillations of $n=3$ becomes wide and their frequencies 
decreases as $c_{\rm A}^2/c_{\rm s}^2$ increases, and finally there is no trapping for 
$c_{\rm A}^2/c_{\rm s}^2 >2$ (figure 5).
This situation is similar to the case where the polytropic index is increased in 
polytropic gases  (see Paper I).

The decrease of frequency of trapped oscillations with increase of $c_{\rm A}^2/c_{\rm s}^2$ 
can be understood by considering the following situations.
Since toroidal magnetic fields are considered here, the nearly vertical oscillations are the 
fast mode among three MHD oscillation modes.
Their frequencies seen from the corotating frame are thus higher than those of 
pure acoustic oscillations in non-magnetized disks.
If we consider purely vertical oscillations, neglecting $u_r$ and $u_\varphi$, in vertically 
isothermal disks with $c_{\rm A}^2/c_{\rm s}^2=$ const., we find easily that the eigen-frequency 
of the oscillations in the corotating frame, $(\omega-m\Omega)$, is given by
\begin{equation}
     (\omega-m\Omega)^2=\biggr[\frac{c_{\rm s}^2+c_{\rm A}^2}{c_{\rm s}^2+c_{\rm A}^2/2}
          (n-1)+1\biggr]\Omega_\bot^2
\label{d1}
\end{equation}
[see equation (\ref{omega})], the right-hand side being larger than $n\Omega_\bot^2$ by the 
presence of toroidal magnetic fields, as expected.
If the terms neglected in deriving equation (\ref{d1}) are taken into account,
$(\omega-m\Omega)^2$ becomes larger than the right-hand side of equation (\ref{d1}).
Hence, we see that one of the propagation region of the nearly vertical oscillation is given by
\begin{equation}
     \omega< m\Omega -\biggr[\frac{c_{\rm s}^2+c_{\rm A}^2}{c_{\rm s}^2+c_{\rm A}^2/2}
          (n-1)+1\biggr]^{1/2} \Omega_\bot.
\label{d2}
\end{equation}
This shows that the upper boundary curve specifying the propagation region of oscillations, i.e.,
$\omega=m\Omega-[(n-1)(c_{\rm s}^2+c_{\rm A}^2)/(c_{\rm s}^2+0.5c_{\rm A}^2)+1]^{1/2}\Omega_\bot$,
moves downwards on the propagation diagram ($\omega$ - $r$ plane) as $c_{\rm A}^2/c_{\rm s}^2$ 
increases (see figures 1 and 2).
In the case of the oscillations of $n=3$, the boundary curve tends to close to $\omega\sim 0$
(as $c_{\rm A}^2/c_{\rm s}^2$ approaches 2.0),
since the term inside the brackets of equation (\ref{d2}) becoms 4.
This is the reason why we have no trapped oscillations for $c_{\rm A}^2/c_{\rm s}^2\simeq 2$
when $n=3$ and $a_*=0$.

Frequencies of QPOs observed in LMXBs have time change, distinct from
those in galactic black hole candidates.
So far as trapped oscillations are concerned, a large change of their frequencies cannot be 
expected, unless magnetic fields are considered.
Global toroidal magnetic fields in accretion disks are time-dependent by amplification due to
winding and by damping due to reconnection.
In the present model of QPOs, the time change of QPOs is attributed to the time change of 
global magnetic fields, but careful discussions are 
necessary whether this is consistent with observations.  

There are some important problems remained to be clarified.
First, it is not clear whether the innermost part of disks can be regarded as a boundary where
oscillations are reflected back outwards.
In disks of standard or ADAF disks, the innermost part of disks will reflect incomming
waves, since the density decreases there sharply inwards.
[See Kato et al. (1988) and Manmoto et al. (1996a,b) for reflection of waves in the inner edge of
disks.]
In slim disks, however, there will be no sharp density decrease inwards near the transonic radius, 
and thus the reflection of waves 
in the innermost region will be not so efficient.
In the case where the central source is a neutron star, the stellar surface or a transition
region near to the surface will, at least, partially reflect incoming waves.

Whether nearly vertical oscillations are really excited on disks is also a problem to be examined,
since they will not be excited by thermal and viscous overstable processes (e.g., Kato 1978),
because of the presence of node(s) in the vertical direction.
Most conceivable processes will be stochastic processes of turbulence (Goldreich and Keely 1977a,b).
In many stars with various characteristics (e.g., different evolutionary stages, effective 
temperatures...), solar-like (non-radial) oscillations have been observed.
The so-called $\kappa$-mechanism cannot excite all of these oscillations.
Their origin is now known to be stockastic processes of turbulence
[see Samadi (2009) for a recent review of stochastic excitation of the oscillations]. 
Compared with in stars, much stronger MHD turbulence by magneto-rotational instability
(MRI) are expected in accretion disks, especially in the inner region of disks.
Hence, it is natural to suppose that in accretion disks many trapped oscillations are
simultaneously excited by turbulence.
This may be one of causes of vareity of QPOs in LMXBs.
This is a problem to be examined in the future.

\bigskip
\leftskip=20pt
\parindent=-20pt
\par
{\bf References}
\par
Blaes, O.M., Arras, P., \& Fragile, P.C. 2006, MNRAS, 369, 1235 \par
Fu, W. \& Lai, D, 2009, ApJ., 690, 1386 \par
Goldreich, P. \& Keely, D.A. 1977a, ApJ, 211, 934 \par
Goldreich, P. \& Keely, D.A. 1977b, ApJ, 212, 243 \par
Kato, S. 1978, MNRAS, 185, 629 \par
Kato, S. 2001, PASJ, 53, 1\par 
Kato, S. 1990, PASJ, 42, 99 \par
Kato, S. 2003, PASJ, 55, 257 \par
Kato, S. 2010, PASJ, 62, 635 (paper I) \par
Kato, S. \& Fukue, J. 1980, PASJ, 32, 377\par
Kato, S., Honma, \& F. Matsumoto, R. 1988, MNRAS, 231, 37 \par
Kato, S., Fukue, J., \& Mineshige, S. 2008, Black-Hole Accretion Disks --- Towards a New paradigm --- 
  (Kyoto: Kyoto University Press)\par
Lai, D. \& Tsang, D. 2009, MNRAS, 393, 979 \par
Latter, H.N. \& Balbus, S.A. 2009, MNRAS, 399, 1058 \par 
Li, L.-X., Goodman, J., \& Narayan, R. 2003, ApJ, 593,980 \par
Manmoto, T., Takeuchi, M., Mineshige, S., Kato, S., \& Matsumoto, R. 1996a, in {\it Physics of
      Accretion Disks}, eds. S.Kato, S.Inagaki, S.Mineshige, J.Fukue (Gordon and Breach, Amsterdam),
      p.57 \par
Manmoto, T., Takeuchi, M., Mineshige, S., Matsumoto, R., \& Negoro, H. 1996b, ApJ, 464, L135 \par
Nowak, M.A \& Wagoner, R.V. 1992, ApJ, 393, 697 \par
Okazaki, A.T., Kato, S., \& Fukue, J. 1987, PASJ, 39, 457\par
Ortega-Rodriguez, M., Silbergleit, A.S., \& Wagoner, R.V. 2002, ApJ, 567, 1043\par
Perez, C.A., Silbergleit, A.S., Wagoner, R.V., \& Lehr, D.E. 1997, ApJ, 476, 589 \par
Samadi, R. 2009, arXiv0912.08175S \par     
Silbergleit, A.S., Wagoner, R., \& Ortega-Rodriguez, M. 2001, ApJ, 548, 335 \par
Tsang, D. \& Lai, D. 2009a, MNRAS, 393, 992 \par 
Tsang, D. \& Lai, D. 2009b, MNRAS, 400, 470 \par
van der Klis, M. 2004, in Compact stellar X-ray sources (Cambridge University Press), 
   eds. W.H.G. Lewin and M. van der Klis (astro-ph/0410551)    \par
Wagoner, R.V. 1999, Phys. Rev. Rep. 311, 259 \par
\bigskip\bigskip

\end{document}